\documentstyle[preprint,prl,aps,epsf]{revtex}
\begin{document}
\draft
\preprint{}
\title{Higher dimensional realizations of activated dynamic scaling 
at random quantum transitions}
\author{T. Senthil and Subir Sachdev}
\address{Department of Physics, Yale University\\
P.O. Box 208120,
New Haven, CT-06520-8120}
\date{\today}
\maketitle

\begin{abstract}
 We show that many of the unusual properties of the one-dimensional random
quantum Ising model are shared also by dilute quantum Ising systems 
in the vicinity of a certain quantum transition in {\em any} dimension $d > 1$. 
Thus, while these properties are not an artifact of $d=1$,  
they do require special circumstances in higher dimensions.

\end{abstract}
\vspace{0.25cm}


\narrowtext

There is considerable current interest in the properties of phase transitions
in random quantum systems. Experimentally accessible quantum transitions, 
such as the transition from an insulator to a metal\cite{MIT} or a 
superconductor\cite{SIT}, and
various magnetic-nonmagnetic transitions in heavy fermion compounds
\cite{HeavyF}, high-$T_c$
cuprates\cite{Subir}, or insulating dipolar Ising magnets\cite{Rosen}, 
often occur in situations with
strong randomness, and are only poorly understood. Theoretically many 
authors\cite{McCoy,Shankar,Fisher} have analyzed the 
random Ising chain in a transverse field -
perhaps the simplest random quantum system. In particular, 
Fisher\cite{Fisher} used a real space renormalization group (RG) approach to
obtain a detailed description of the thermodynamics and static correlation
functions in the vicinity of the critical point. The properties of the system 
were found to be very unusual as compared to conventional quantum critical 
points. Specifically, length scales were found to diverge logarithmically
with energy scales, implying a dynamic critical exponent $z = \infty$. 
The transition was shown to be flanked on either side by ``Griffiths'' 
regions with a susceptibility diverging due to contributions from statistically
rare fluctuations. There are very few reliable results on other random quantum
transitions, especially in finite dimensions $d > 1$; 
thus it is important to understand if the anomalous properties of the random
quantum Ising chain are a specialty of $d = 1$, or if there exist higher
dimensional quantum transitions which share these properties. Numerical
work on higher dimensional transverse field Ising spin glasses has
found evidence for the presence of Griffiths regions, but the dynamic
scaling at the critical point seems conventional\cite{Muyu}.

\begin{figure}
\epsfxsize=5in
\centerline{\epsffile{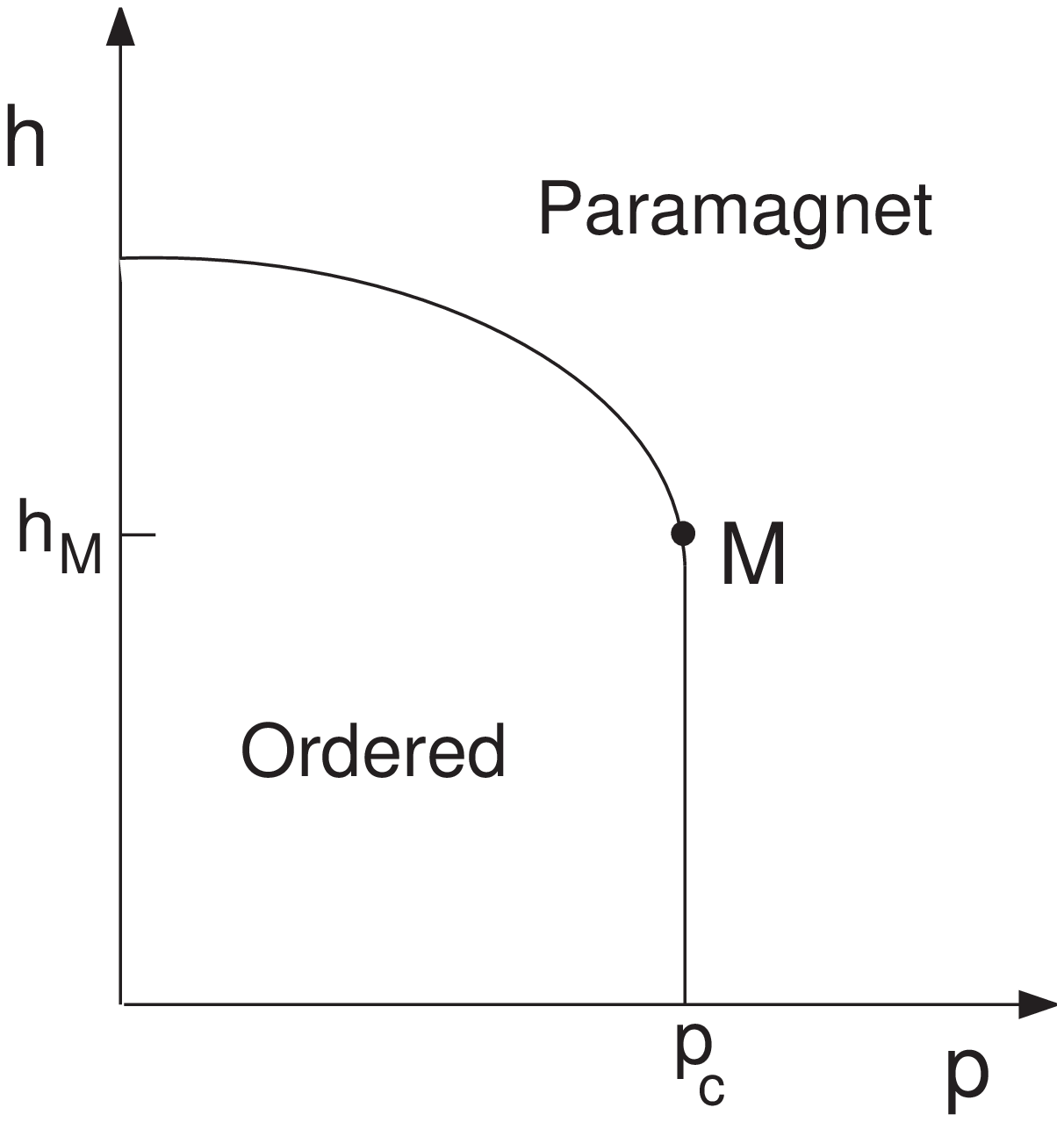}}
\vspace{0.5in}
\caption{Phase diagram of the dilute Ising model in a transverse field ($h$)
at zero temperature. The dilution probability is $p$. The multicritical
point $M$ is at $p=p_c$, $h=h_M$.
The quantum transition along the vertical phase boundary ($h<h_M$, $p=p_c$)
is controlled by the classical percolation fixed point at $p=p_c$, $h=0$;
quantum effects (due to a non-zero $h$) are dangerously irrelevant, and lead to
activated dynamic scaling near the $h<h_M$, $p=p_c$ line. }  
\label{fig1}
\end{figure}

In this paper, we provide a simple example of such anomalous scaling in higher
dimensional random quantum Ising systems. We consider bond or site diluted Ising
models with short range interactions. As was first suggested by Harris~\cite{Harris},
and as we argue below, these models have two
quantum transitions (See Fig~\ref{fig1}): At low dilution, below the
percolation threshold, there is a phase transition when the long range ferromagnetic
order is destroyed by increasing the transverse field. This is expected to be in the 
universality class of the generic random bond quantum Ising transition. Right at
the percolation threshold, there is a finite range of transverse field strengths
at which the system remains critical. There is thus another quantum transition,
across the percolation threshold, at low but non-zero transverse field strengths,
which is potentially in a different universality class.
 These two critical lines meet 
at a multicritical point ($M$). The properties of the second transition, at the
percolation threshold, 
are determined largely by the statistics and geometry of the percolating 
clusters about which much information is available. This permits us to make
definitive statements about the scaling properties of this transition in any
dimension. We show that the dynamic scaling is activated, with length scales 
diverging logarithmically with energy scales. We also demonstrate the existence
of Griffiths phases on either side with diverging susceptibility. 
Our approach shows clearly the connection between these behaviors, and $T=0$
phase transitions at which quantum fluctuations are ``dangerously
irrelevant''~\cite{Fisher,sro}; indeed, various 
exponents of the transition are given by those of the 
classical percolation transition, and are hence known either exactly or
numerically. We also obtain bounds on exponents characterizing the 
multicritical point.

For concreteness, we consider 
{\em bond}-diluted Ising models
defined by the Hamiltonian
\begin{equation}
\label{Ham}
{\cal H} = -\sum_{\langle ij\rangle}J_{ij}\sigma_i^z \sigma_j^z - \sum_i h
\sigma_i^x
\end{equation}
where $\langle ij\rangle$ labels the nearest-neighbor
sites of a $d$-dimensional lattice ($d > 1$), and
$J_{ij}$ equals 0 with probability $p$ and equals $J > 0$,  
with probability $1 - p$. The transverse field $h$
is taken to be non-random, although our results remain valid for weakly random
distributions as well. At $p = 0$, as $h$ is increased, there
is a zero temperature phase transition from an ordered ground state
to a disordered one (See Fig~\ref{fig1}). 
On the other axis, when $h = 0$, so that the
system  is classical, there is a percolation transition at $p = p_c$. For 
$p < p_c$,
there is a thermodynamically large connected cluster, while for $p > p_c$,
there only are finite connected clusters. For
$p < p_c$, for small enough $h$, the system retains long range order. This
is ultimately destroyed for some $h > h_c(p)$, with $h_c (p)$ expected to be
a  monotonically decreasing function of $p$. On the other hand, if $p > p_c$,
there is no long range order for any $h$. 

Now consider $p = p_c$. Again 
there is zero magnetization and no long range order for any $h$. However 
the system stays critical for $h < h_M = h_c(p_c)$ (Fig~\ref{fig1}). To see this, note
that, although there is no thermodynamically large connected cluster at $p_c$, there
still is an infinite connected cluster with a fractal dimension smaller
than the spatial dimension. The spins on this cluster align together at 
$h = 0$. A small but non-zero $h$ is not sufficient to destroy this order
on the critical cluster. In fact, two spins on any sufficiently large finite
cluster remain strongly correlated with each other for small $h$. The critical 
cluster eventually loses order when $h$ reaches $h_M$. To make this more 
precise consider the disorder averaged, equal time ($\tau$), two point spin correlation
function. Spins  at points $0$ and $x$ are correlated only if they belong to the same
cluster; thus 
$G(x) \equiv \left[ \langle \sigma^z (x, \tau=0) \sigma^z(0,0) \rangle
- \langle \sigma^z (x,0) \rangle \langle \sigma^z (0,0) \rangle \right]
 = \int dC C 
{\cal P} $($0$ and $x$ belong to the 
same cluster) ${\cal P}$(if $0$ and $x$ belong to the same cluster, they have
correlation $C$), where angular brackets are averages over quantum/thermal
fluctuations, square brackets represent disorder averages, and ${\cal P} (E)$
is the probability of the event $E$. At
$p = p_c$, the first probability 
$\sim x^{-d + 2 - \eta_p}$ for large $x$ \cite{Stauffer}. The 
arguments above imply that the integral over $C$ is non-zero and independent
of $x$ for large $x$ for $h < h_M$. Thus $G(x, p = p_c) \sim   
x^{-d + 2 - \eta_p}$ for large $x$ for $h < h_M$, and there is a critical
line for $h < h_M$ at $p = p_c$.

In this paper, we will primarily focus attention on the transition
across this critical line. We show that at $T = 0$, the properties of this
transition are strikingly similar to the $d = 1$ results for the random
quantum Ising chain \cite{Fisher}. First consider the 
equal-time two point function away 
from $p_c$. Arguing as in the previous paragraph, we get 
$G(x)
\sim {\cal P}$($0$ and $x$ belong to the same cluster) 
$\sim x^{-d + 2 - \eta_p} f(x/\xi)$ for large $x$ where the
correlation length $\xi \sim |p - p_c|^{- \nu_p}$. Similarly, for $p < p_c$
the mean 
magnetization $ \left[ \langle \sigma^z (x, 0) \rangle \right] \sim {\cal P}$(any
given site belongs to infinite cluster)
$\simeq (p_c - p)^{\beta_p}$. The exponents $\beta_p, \eta_p, \nu_p$ are those of
classical percolation theory, and
are known exactly in $d = 2$ and for $d > 6$. For any $d < 6$, they satisfy
 $\beta_p = \nu_p(d - 2 + \eta_p)/2$.

 Now consider
dynamic correlations. The imaginary part of the local dynamic susceptibility
$\left[ \chi_L''(\omega) \right] = \Sigma_N {\cal P}$(given site $i$ belongs to 
cluster of $N$ sites)$\int y dy {\cal P}$(if site  $i$ belongs to cluster
of $N$ sites, $\chi_i''(\omega) = y$). The energy levels of a cluster
of $N$ sites can be described for $h \ll J$ as follows: The two lowest levels
correspond to the states of a single effective Ising spin with
magnetic moment $\sim N$ in an effective 
transverse field $h_{{\rm eff},N}$. For large $N$, $h_{{\rm eff},N}$ can be estimated
in $N^{th}$ order perturbation theory to be $\tilde{h}\exp(-c N)$. (In
general, there would also be a prefactor that varies as a power of $N$; we will
ignore this as it is subdominant to the exponential in $N$. When necessary, 
we will indicate the modifications induced by keeping this prefactor).
The
quantities $\tilde{h}$ and $c$ are of order $h$ and $\ln({J}/{h})$ 
respectively but
their precise values depend on the particular cluster being considered.  As
the distribution of $\tilde{h}$ and $c$ is not expected to
become very broad near the transition\cite{note}, we will replace them by their
typical values $h_0$ and $c_0$ respectively. Apart from these two lowest
levels, there are other levels separated from these by energies $\sim J$.
These can be ignored for the low energy physics, and for small $\omega \ll h$, we
only need to consider large clusters. 

We now need the
following results\cite{Stauffer} from percolation theory for the probability $P(N,p)$
that a given site belongs to a large cluster of $N$ sites. At $p = p_c$,
$P(N,p_c) \sim N^{1-\tau}$
where $\tau$ equals $5/2$ for $d > 6$, and equals $1 + d/D$ for 
$d < 6$ where $D$ is the fractal dimension 
of the critical clusters. In particular $\tau = 187/91$ in $d = 2$,
and is approximately $2.18$ in $d = 3$.
Away from $p_c$, for $d < 6$ or $d > 8$, $P(N, p)$ satisfies the
scaling form
\begin{equation}
P(N, p) \sim N^{1-\tau}g\left( N/\xi^D \right)
\end{equation}
The scaling function $g(y)$ is universal but is different for $p < p_c$ and 
$p > p_c$; it approaches $1$ for $y \ll 1$, while for $y \gg 1$,  
\begin{eqnarray}
g(y, p > p_c) & \sim & y^{-\theta + \tau} e^{-c_{+}y}  \nonumber \\
g(y, p < p_c) & \sim & y^{-\theta' + \tau} e^{-c_{-}y^{1 - 1/d}}
\end{eqnarray}
with $\theta = 1$, ${3}/{2}$ and ${5}/{2}$ for $d = 2, 3$
and $d > 8$ respectively, and $\theta' = 
{5}/{4}$, $-{1}/{9}$ in $d = 2, 3$ respectively. The constants 
$c_{+,-}$ are
of order unity.
For $6 < d < 8$, $P(N,p)$ satisfies a more complicated two-variable scaling
form~\cite{Lub}. For simplicity, we shall not discuss this case here, 
though including it is straightforward.

Using these percolation results, we get for the dynamic susceptibility of the Ising
model
\begin{eqnarray}
\left[ \chi_L''(\omega) \right] & \sim & \int \frac{dN}{N^{\tau - 1}} g(N/\xi^D)
\delta(\omega - h_0 e^{- c_0 N})  \nonumber \\ 
& \sim & \frac{1}{\omega(\ln(h_0/\omega))^{\tau - 1}} 
g \left(\frac{\ln(h_0/\omega)}{c_0 \xi^D}\right)  
\end{eqnarray}
Note that the scaling variable is $\ln(1/\omega)/{\xi^D}$.
This is a precise statement of the activated dynamic scaling mentioned
earlier, which has thus been shown to occur in all $d > 1$ in the present model.
Asymptotic forms in various limits can be obtained from the
limiting behavior of $g(x)$ described above. For $p \geq p_c$, we get
$\chi_L'' \sim \omega^{-1 + \alpha} (\ln(h_0/\omega))^{1-\tau}$
with $\alpha \sim \xi^{-D}$ so that at $p = p_c, \alpha = 0$ 
(Including the power-law prefactor in $h_{{\rm eff},N}$ will only change the 
power of $\ln(1/\omega)$ in by $\sim \xi^{-D}$, and similarly
for the prefactor in the expression for $p < p_c$ given below).
Note that just on the disordered side, the system is gapless with a
power-law density of states. The physical origin of this is, as usual,
the presence of rare, large clusters with arbitrarily small energy gaps.
For $p < p_c$, the presence of the infinite cluster (and the associated long
range order) gives rise to a delta function at $\omega = 0$. 
For $\omega \neq 0$, $\chi_L''(\omega)$ is still determined by contributions from
the finite clusters. Proceeding as before, we find $\chi_L''(\omega \neq 0)
\sim (1/\omega) (\ln(h_0/\omega))^{1-\tau}
\exp\left(-\kappa (\ln(h_0/\omega))^{1 - 1/d}\right)$ with 
$\kappa \sim \xi^{-D(1 - 1/d)}$. Again the system is gapless. The
gaplessness of both the ordered and disordered phases in the vicinity of
the transition is unlike quantum transitions in pure systems, but is 
probably generic to many random quantum transitions\cite{Fisher,Thill,Muyu}.

 The magnetization in response to a uniform external applied magnetic
field $H$ along the $\hat z$ direction can be calculated similarly. For
small $H \ll h$, only large clusters contribute. The magnetization per site of
a cluster of size $N$ is that of an Ising spin of magnetic moment $N$
in a transverse field $h_{{\rm eff},N}$, and is therefore given by
\begin{displaymath}
M_N(H) = \frac{NH}{((NH)^2 + h_{{\rm eff},N}^2)^{1/2}}
\end{displaymath}
Thus the total magnetization per site (after subtracting the regular contribution
of the infinite cluster for $p < p_c$) is
\begin{displaymath}
M(H) - M(H = 0)  \sim \int dN \frac{1}{N^{\tau - 1}}g(N/\xi^D) M_N(H) 
\end{displaymath}
The singular part therefore has the scaling form
\begin{equation}
M_{{\rm sing}}(H) \sim \frac{1}{(\ln(h_0/H))^{\tau - 2}}\Phi\left(c \frac{\ln(h_0/H)}
{\xi^D}\right)
\end{equation}
with $c$ a non-universal constant, and $\Phi(y)$ a universal function which is
related to 
$g(y)$ by
\begin{displaymath}
\Phi(y) = \int_{1}^{\infty} w^{1-\tau} dw g(wy) 
\end{displaymath}
Again note that the scaling variable is $\ln(h_0/H) \xi^{-D}$, which is
unlike conventional critical behavior, but is similar to the $d = 1$ random
quantum Ising chain~\cite{Fisher}.

We thus have the following asymptotic forms as $H \rightarrow 0$:
\begin{equation}
M_{{\rm sing}}(H) \sim \left\{
\begin{array}{cc}
  (\ln(h_0/H))^{2-\tau} & p = p_c \\
 \xi^D (\ln(h_0/H))^{1-\theta}({H}/{h_0})^{\gamma} & p \geq p_c \\
\xi^{D(1 - 1/d)} (\ln(h_0/H))^{-\theta' + 1 - 1/d}
e^{- (\gamma' \ln(h_0/H))^{1 - 1/d}} & p \leq p_c
\end{array}
\right.
\end{equation}
with $\gamma, \gamma' \sim \xi^{-D}$ (as with the dynamic susceptibility,
including the power-law prefactor in $h_{{\rm eff},N}$ changes the power of 
$\ln({1}/{H})$ for $p>p_c$, and the prefactor for $p< p_c$
by $\sim \xi^{-D}$).    
Note that the magnetization rises as a power of $H$, with a continuously varying
exponent which is smaller than $1$ in a region of the disordered phase
close to the transition. This is again similar to the $d = 1$ result, and gives 
rise to a divergent linear susceptibility throughout this region. In the ordered
side ${dM}/{dH} \sim {1}/{H}$ with weak corrections. Thus the linear
susceptibility diverges in the ordered side as well. 

A further similarity with the $d = 1$ results can be found by 
generalizing the problem to systems with Potts symmetry. For the
random $q$-state quantum Potts model in $d = 1$, it has been shown\cite{Sen-Sat}
that all the critical exponents and the scaling functions for the 
mean spatial correlations and magnetization are independent of $q$, {\em i.e},
they are the same as for the Ising case. For the diluted models that we
have considered here, it should be clear that a vertical critical line exists at
the percolation threshold for all $q$. The exponents and appropriate
scaling functions of the corresponding
quantum transitions are again independent of $q$ as they are determined
mainly by the geometric properties of the lattice near the percolation
threshold. As in $d = 1$, all the $q$ dependence is in non-universal quantities
and in a high-energy cutoff limiting the regime of universal scaling behavior.

\begin{figure}
\epsfxsize=5in
\centerline{\epsffile{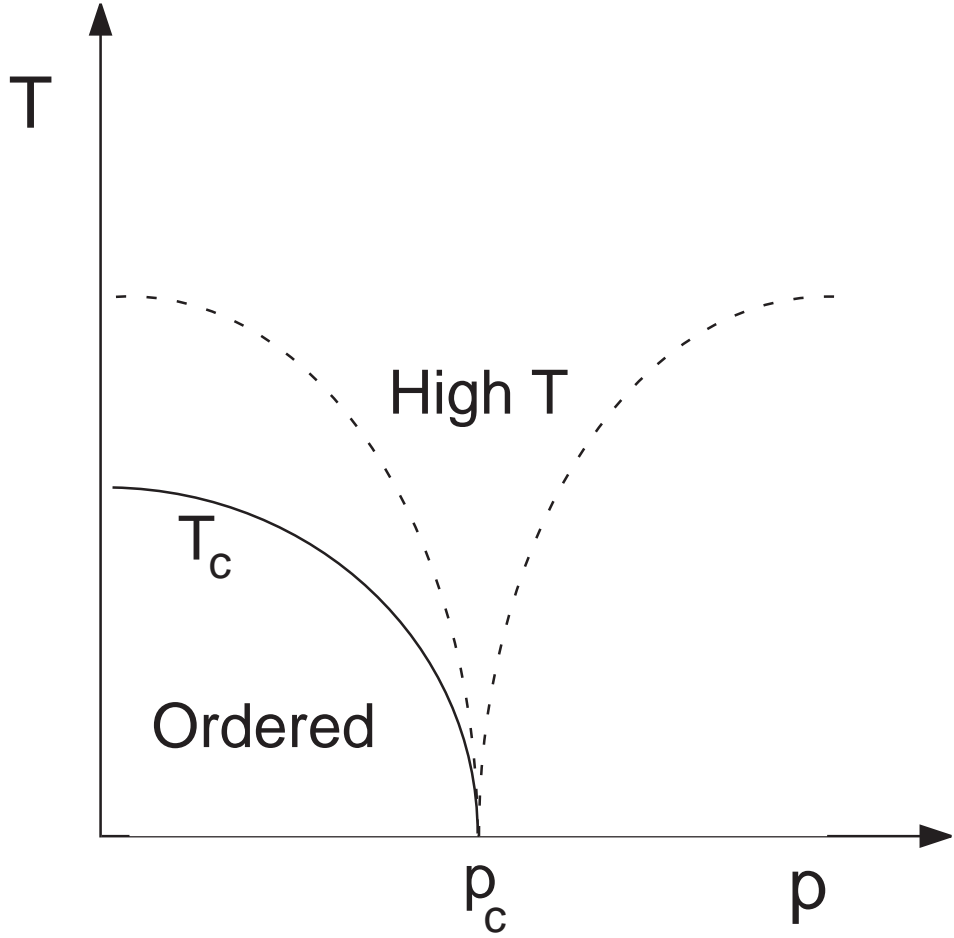}}
\vspace{0.5in}
\caption{Finite temperature phase diagram for $h < h_M$.
The dashed lines ($T \sim 1/\ln(1/|p-p_c|)$) represent crossovers
from the high $T$ regime, characterized by spin fluctuations on the critical
infinite cluster, to the low $T$ regimes.
The solid line for $p < p_c$ is the
phase transition ($T=T_c \sim 1/\ln(1/(p_c-p))$ where long range order is
destroyed. 
}  
\label{fig2}
\end{figure}

So far, we have focussed on the $T = 0$ scaling properties, and demonstrated 
the similarities with the $d = 1$ results. What about the finite $T$ 
properties ? 
For the {\em classical} dilute Ising model
at $p=p_c$, the correlation length at finite $T$
behaves as $\exp(\mbox{contant}/T)$~\cite{Harris}. This is essentially due to the
presence  of one dimensional segments in the critical percolating clusters. For
the  quantum problem for $h < h_M$, these one dimensional segments would
give rise to a thermal correlation length ($\xi_T$) with a similar exponential
dependence on $1/T$, and
a prefactor
that is a power-law in $T$; this is the behavior in the ``high $T$'' region
of Fig~\ref{fig2}. In
contrast, for the
$d = 1$ random quantum Ising chain, the correlation length rises as a power of
$\ln(1/T)$ at the critical couplings. 
Away from the critical point, the crossovers are as shown in Fig~\ref{fig2}.
The low $T$ behavior appears when $\xi \sim \xi_T$, or $T \sim \ln^{-1}
(1/(|p-p_c|)$. On the disordered side, the low $T$  system is described well as
a collection of rigid Ising clusters with  effective transverse
fields and a size distribution as before; this leads, for instance,
to a linear susceptibility $\chi_T \sim T^{-1 + \kappa}$ (up to
$\log$ corrections) with $\kappa \sim \xi^{-D}$.
On the ordered side, there is a 
finite temperature phase transition; as in the classical case,
as $p \nearrow p_{c}$, the transition temperature falls to zero
as $T_c \sim \ln^{-1} (1/(p_c - p))$. 

We now turn to the special point $M$ (Fig~\ref{fig1}).
Correlations at the point $M$ should
decay faster than along the vertical critical line considered
above. Thus, if $G_{M}(x) \sim x^{-\phi}$, then
$\phi > d - 2 + \eta_p$. Similarly, as $p$ approaches $p_c$
from below at $h_M$, the magnetization should go to zero
faster than for $h < h_M$. Thus $\beta_M > \beta_p$. A detailed
theory of the scaling at this point we leave as an open 
question.

Finally it is interesting to ask if this vertical critical line at the 
percolation threshold continues to exist for systems with continuous
symmetry, such as the $O(N)$ rotor models. For $N > 2$, the presence of one
dimensional segments in the critical clusters implies rapidly decaying 
correlations in large clusters. Thus the vertical critical line will not
be present for $O(N)$ models for $N > 2$. The $O(2)$ case is special as
correlations only decay as a power-law in one dimension. Whether this is 
like the Ising or the $O(N > 2)$ system we leave as another open 
question.

To summarize, we have presented a simple example of a 
random quantum transition in dimensions $d > 1$ 
which exhibits many of the 
properties of the transition in the $d = 1$ random quantum 
Ising chain. In particular, the dynamic scaling was activated,
with $\ln(1/\mbox{energy scale}) \sim \xi^D$. Further, there were Griffiths
regions on either side of the transition, with a singular
density of states and a diverging susceptibility. It would be 
interesting to find experimental systems where this transition can be studied.
(One possibility is the system $LiHo_xY_{1 - x}F_4$ where, however, 
the presence of
dipolar interactions may complicate the situation.) 
Theoretically, an important feature
of this transition, as in the $d = 1$ system, is that it is controlled by
a classical fixed point with quantum fluctuations being ``dangerously
irrelevant". 
This feature is also found in quantum Ising models
in a random longitudinal field, which undergo, for $d > 2$, a phase transition
from an ordered to a paramagnetic phase at $T = 0$~\cite{Aharony}. 
In fact, this fixed point
also controls the non-zero $T$ phase transition, where it has been argued
that the classical dynamic scaling is 
activated~\cite{Villain}. Extending the argument to the 
quantum dynamics at the $T = 0$ transition, suggests that the quantum dynamic
scaling would again be activated\cite{qrf}. Further, classical fixed
points have also appeared in recent studies of the metal-insulator
transition~\cite{bk}, and in the $T=0$ onset of spin-glass order in
metallic systems~\cite{sro}; whether, in these cases, the dynamic scaling is
activated or conventional, due to the presence of itinerant fermions, is an
interesting question, worthy of future investigation.

     One of us (T.S) would like to thank S.N.~Majumdar for a 
collaboration on some related earlier work\cite{Sen-Sat}. We
also thank Y.~Gefen, N.~Read, and T.F.~Rosenbaum for useful discussions.
This research was supported by NSF Grant No DMR 96-23181.

\begin{references}
\bibitem{note} In particular, we expect the distribution of $c$ to be bounded.
\end{references}
\end{document}